# Investigations of the Effects of Pressure on the Structural and Electronic Properties of Co$_2$VZ (Z = Al, Be) Full Heusler Alloy: A Comparative Study Using DFT

Sumit Kumar[1,2*], Diwaker [3†], Shyam Lal Gupta[4*], Karan Singh Vinayak[1]

[1*]Department of Physics, DAV College , Sec-10, Chandigarh, 100190, India.
[2]Department of Physics, Government College, Una, City, 174303, Himachal Pradesh, India.
[3]Department of Physics, Rajiv Gandhi memorial Govt. College, , Jogindernagar, 175015, Himachal Pradesh, India.
[4]Department, HarishChandra Research Institute(HRI), , Prayagraj, 211019, U.P, India.

*Corresponding author(s). E-mail(s): sumitphysics@gmail.com;
shyamlalgupta@hri.res.in;
Contributing authors: diwakerphysics@gmail.com;
drksvinayak.dav@gmail.com;
†These authors contributed equally to this work.

## Abstract

This study focuses on the investigations and comparative study of the electronic structure of Co$_2$VZ (Z = Al, Be) Heusler alloys under varying high pressure conditions. The pressure range explored spans from 0.0 GPa to 30.0 GPa, with increments of 0.5 GPa. The WIEN2K simulation programme is used to investigate the effect of pressure on the structural, magnetic, and electronic properties of Co$_2$VZ (Z = Al, Be) Heusler alloys. The WIEN2K simulation code's WC-GGA and mBJ exchange correlation potentials are used to investigate various features. The results of the very first investigation using the WC-GGA exchange correlation potentials are then compared to earlier experimental and theoretical findings employed different exchange correlation potentials.The stability observed in the P-V plot indicates the absence of any structural phase transition from a cubic symmetry structure to another structural phase. The varying slopes observed in the band gap's response to increasing pressure in different pressure ranges for Co$_2$VZ (Z = Al, Be) alloys can be attributed to the predominance of either permittivity or quantum confinement effects

**Keywords:** Heusler alloys, Band structure, DOS, DFT

# 1 Introduction

Over the past few decades, intermetallic alloys have garnered considerable attention in the advancement of memory devices, magneto-electronics, and spintronics. These include magneto-resistive random access memories, spin transistors, spin torque oscillators, and spin valve generators [1–4]. Achieving high spin-polarized current is crucial for optimal performance of spintronic devices. Cobalt-based half and full Heusler alloys are particularly important for generating full spin-polarized current [5–7]. The concept of spin polarization is based on the presence of metallic characteristics in one spin channel (up or down) and semiconductor behavior in the other spin channel (up or down). Hence, half-metallic Heusler alloys play a crucial role in this context [8, 9]. The ease of synthesis due to their high Curie temperature makes Heusler alloys highly desirable. Full-Heusler alloys of the X$_2$YZ type, possessing a face-centered cubic structure, have been extensively studied worldwide with different element substitutions at atomic positions X, Y, and Z [10–16]. This broad and comprehensive research suggests ample opportunities for exploring the properties of Heusler alloys through the substitution of various elements for different applications. Apart from full Heusler alloys, there are two other types of Heusler alloys: half-Heusler alloys with the formula XYZ and quaternary Heusler alloys with the formula XXYZ. In all these types, X, X$_2$, and Y are transition metals from the 3d, 4d, and 5d groups, while Z is a non-magnetic element from the 3d, 4th, and 5th groups of the periodic table [17–20]. Heusler alloys such as Ti$_2$CoB, Ni$_2$MnB, and Co$_2$NbB have been investigated to explore their properties under pressure due to their hardness, high melting point, and resistance to oxidation [21–23]. The stability of Heusler alloys at high temperatures makes them well-suited for high-temperature power generation and shape memory devices [24, 25]. In recent years, research on cobalt-based Heusler alloys has significantly increased, particularly due to their high tunnel magnetoresistance [26]. Rai et al. conducted an extensive study on the electronic, magnetic, and optical properties of cobalt-based Heusler alloys, highlighting their applications in spintronics [27]. Numerous other cobalt-based Heusler alloys, such as Co$_2$MnSi [28], CoMnTiAl [29], CoMnVAs [30], Co$_2$MnZ (Z = Ge, Sn) [31], CoVTiAl [32], Co$_2$FeZ (Z = Al, Ga, Ge, S) [33], CoCuMnZ (Z = In, Sn, Sb) [34], CoZrIrSi [35], NbVMnAl, and NbFeCrAl [36], CoFeCrZ (Z = P, As, Sb) [37], FeRhCrZ (Z = Si and Ge) [38], and YFeCrZ (Z = Al, Sb, Sn) [39], have been investigated for their intriguing properties and reported as half-metallic alloys. Guezlane *et al.* calculated the electronic structure of Co$_2$Cr$x$Fe$1 − x$X (X = Al, Si) and identified it as a full-Heusler ternary compound with half-metallic characteristics [40].T. kanomata

*et.al.* investigated the magnetic characteristics of $Co_2VAl$ and $Co_2VGa$ under pressure experimentally and computationally [51]. While the previous investigations were conducted at atmospheric pressure, it is crucial to highlight that there are few publications in the literature on the effect of pressure on the physical properties of Heusler alloys. In this study, we report for the first time on pressure investigations of $Co_2VBe$, while experimental and theoretical results of previously studied $Co_2VBAl$ are contrasted with first-time reporting of the compounds via wcGGA and mBJ xc potentials. Pressure is important in defining the physical properties of materials, and investigating material deformation under pressure has grown increasingly interesting. Such investigations provide valuable insights into the behavior of materials by observing fundamental parameters. Despite the fact that literature on numerous types of Heusler alloys, including understudied alloys, is accessible, the increasing demand for spintronic devices needs a comparative analysis of current alloys to bridge the gap between experimental and theoretical outcomes. Under changing hydrostatic pressures ranging from 0 to 30 GPa, we investigate the structural, magnetic, and electronic properties of experimentally and theoretically researched $Co_2VAl$ and experimentally unexplored $Co_2VBe$ Heusler alloys. We researched alloys with distinct exchange correlation potentials that were not previously studied to the best of our ability. The findings of the study are discussed in depth in the following sections.

## 2 Computational Details

The properties of $Co_2VZ$ (Z = Al, Be) Heusler alloys were investigated using the FP-LAPW method within the density functional theory (DFT) framework. The WIEN2k simulation package was employed for these investigations [41]. Different approximations were used to handle the exchange and correlation potentials. In the GGA scheme, the band profile tends to be underestimated due to imperfect handling of strongly correlated mechanisms. To address this issue and improve accuracy, a modified Becke-Johnson (mBJ) potential was employed. The exchange-correlation potential used in the calculations can be expressed as:

$$V_{X,\sigma}^{mBJ}(r) = CV_{X,\sigma}^{mBJ}(r) + (3C-2)\frac{1}{\pi}\sqrt{\frac{5}{12}}\sqrt{\frac{2t_\sigma(r)}{\rho_\sigma(r)}} \quad (1)$$

where $t_{\sigma(r)} = \frac{1}{2}\sum_i^{N_\sigma} \nabla_{\psi_{i,\sigma}}$ represents the kinetic energy and $\rho_{\sigma(r)} = \sum_i^{N_\sigma} \psi_i^{N_\sigma}.\psi_{i,\sigma}$ as the charge desnity. The conventional unit cell is divided into non-overlapping spheres with a spherical wave function and an interstitial region with a plane-wave character. The energy convergence limit was set to 0.0001 Ry, and the plane wave expansion used a cutoff of RMT×$K_{MAX}$, where $K_{MAX}$ is the cutoff for the wave function basis and RMT is the smallest radius of the muffin-tin spheres. Integration of the Brillouin zones employed a dense K-mesh of approximately 1000 k points, and a cutoff energy around -6 Rydberg was chosen to separate the valence and core states. The minimum selected values for RMT radii were Al = 2.06, V = 2.17, and Co = 2.28 for $Co_2VAl$, and Be = 1.81, V = 2.07, and Co = 2.18 for $Co_2VBe$.

| Table 1-: Calculated a($A^0$), B(GPa), $B'$, ($E_0$) for $Co_2VZ$(Z=Al,Be) | | |
| --- | --- | --- |
| Parameter | $Co_2VAl$ | $Co_2VBe$ |
| Lattice constant a ($A^0$) | 5.699 (Present work) | 5.455 |
| | 5.760 [42] | 5.51 [42] |
| | 5.759 [43] | |
| | 5.800 [44] | |
| | 5.754 [45] | |
| | 5.770 [46] | |
| Bulk modulus in (Gpa) | 216.3848 (Present work) | 225.7989 (Present work) |
| | 199.748 [42] | 216.318 [42] |
| | 197.740 [43] | |
| Pressure Derivative of Bulk Modulus($B'$) | 4.85(Present work) | 4.1577(Present work) |
| | 4.35 [42] | 4.60 [42] |
| $V_0$ | 312.3408 (Present work) | 273.9062 (Present work) |
| | 322.544 [42] | 282.597 [42] |
| ($E_0$) Ry | -7955.6203 (Present work) | –7499.7929 |
| | -7958.312 [42] | - |
| Formation energy | -1.81221 (Present work) | -1.79553 |
| | -1.80521 [42] | -1.79217 [42] |

# 3 Structural properties

The structural properties of the investigated Heusler alloys are of great importance and are detailed as follows. The unit cell configurations for both Heusler alloys, $Co_2$VAl and $Co_2$VBe, are visually presented in Figure 1 and Figure 3, respectively. In order to optimize the unit cell, calculations were performed for both the non-magnetic (NM) and spin-polarized (SP) phases. The optimization curves obtained during this process are depicted in Figure 2 and Figure 4, respectively.

Upon analyzing the optimization curves, it becomes evident that the spin-polarized phase exhibits greater stability for both unit cells. This finding highlights the influence of magnetism on the structural properties of the Heusler alloys and underscores the importance of considering spin polarization when studying these materials.

The optimized lattice parameters obtained from the optimization process serve as crucial inputs for the calculation of electronic and thermal properties. These lattice parameters represent the most energetically favorable configurations of the unit cell.

To assess the stability of the compounds, the formation energy of the Heusler alloy is computed using the following formula:

$$E_{FE} = E_{Co_2VZ} - 2E_{Co_2} - E_V - E_Z \qquad (2)$$

Here, $E_{Co_2VZ}$ represents the calculated equilibrium total energy of the Co2VZ alloy (where Z can be either Al or Be), while $ECo_2$, $E_V$, and $E_Z$ denote the energies per atom of pure Co, V, and Z (Al or Be) in their respective bulk structures.

The calculated values of various structural properties are summarized in Table 1. These properties include the lattice constant $a$ in angstroms ($A°$), bulk modulus $B$ in gigapascals (GPa), pressure derivative of the bulk modulus $B'$, equilibrium volume, ground state energies $E_0$, and formation energies. These properties provide valuable insights into the stability, bonding, and energetics of the Heusler alloys under investigation.

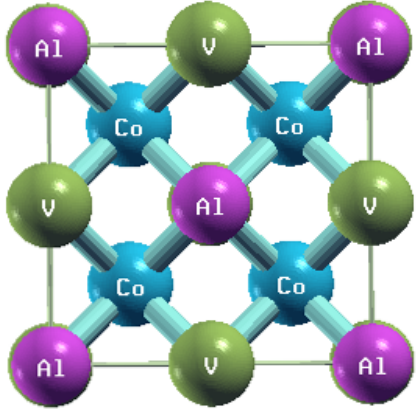

**Fig. 1** Unit Cell of $Co_2VAl$ Heusler alloy

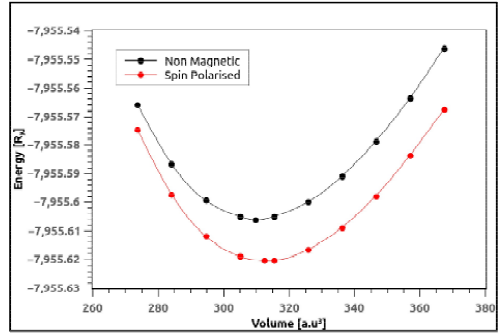

**Fig. 2** Total Energy Vs Volume Curves for $Co_2VAl$ optimization

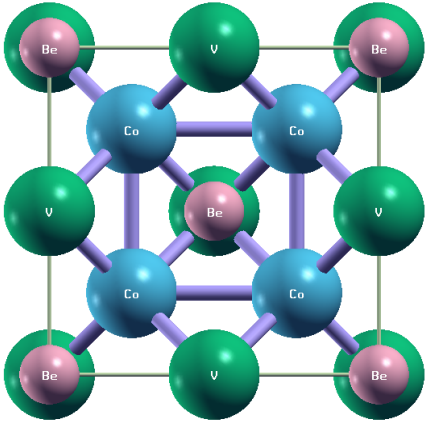

**Fig. 3** Unit Cell of $Co_2VBe$ Heusler alloy

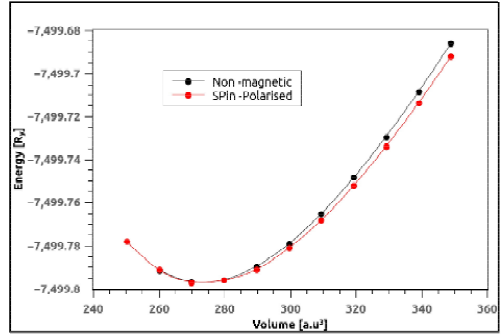

**Fig. 4** Total Energy Vs Volume Curves for $Co_2VBe$ optimization

## 4 Electronic and Magnetic properties

### 4.1 Electronic Properties of $Co_2VZ$ (Z=Al, Be)

Electronic properties play a crucial role in various memory, smart, and spintronic devices, as they are closely tied to the energy band gap and electronic structure of materials. In this study, we focus on investigating the electronic properties of a specific material, namely the Heusler alloy. To characterize the electron structure of this material, we analyze the band structure and total and partial densities of states. The spin-polarized band structures for both spin-up and spin-down channels are examined using WC-GGA and mBJ approximations, as depicted in Figures 5–12.

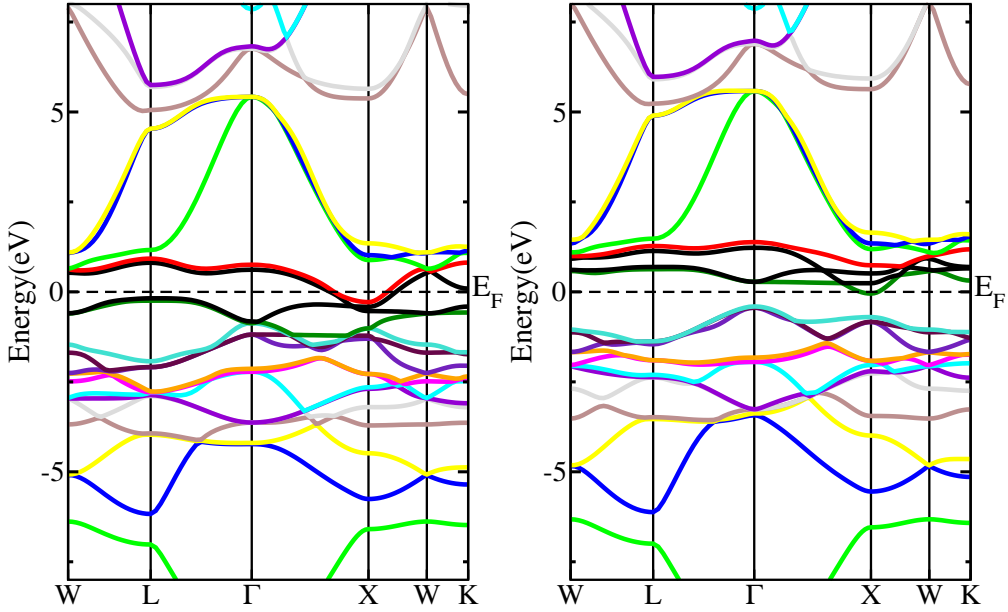

**Fig. 5** Calculated band structure of for $Co_2VAl$ (spin up channel) using WCGGA

**Fig. 6** Calculated band structure of for $Co_2VAl$ (spin down channel) using WCGGA

Analyzing the energy band structures, we observe that the spin-up channel of the $Co_2VAl$ Heusler alloy does not exhibit a band gap. However, a band gap of 0.6 eV is observed in the spin-down channel for the direct transition $E_{\Gamma-\Gamma}$, and 0.22 eV for the indirect transition $E_{\Gamma-X}$. Additionally, we calculate the spin-polarized densities of states for these Heusler alloys, which are presented in Figures 13–16. These figures provide a visual representation of the total density of states for both spin-down and spin-up channels, exhibiting consistency with the energy band structures. Figures 5 and 6 showcase the total DOS for the spin-down and spin-up channels specifically for the $Co_2VAl$ Heusler alloy, which comprise the energy band structures. Additionally, we calculated the total and local magnetic moments of the $Co_2VAl$ Heusler alloy, and these results are summarized in Table 3.

## 4.2 Magnetic Properties of $Co_2VZ$ (Z=Al, Be)

The calculated atomic and total magnetic moments for the compounds under study are calculated with both WCGGA and PBE-GGA exchange correlation potentials. The obtained results are compared with previously published findings using different approximations and are summarized in Table 2. The expected Magnetic moment for $Co_2VAl$ and $Co_2VBe$ is 2 and 1 respectively according to Slater -Pauling rule. The experimental value of magnetic moment for $Co_2VAl$ is 1.86. The difference in experimental and theoretically predicted value of magnetic moment of $Co_2VAl$ is owing to the precipitation of the bcc particles in the matrix L21 phase [51]. it indicates that the magnetic behavior of the alloys under study follows the expected trend based on the number of valence electrons involved. In other words, the magnetic properties of the alloys under study can be reasonably predicted by considering the valence electron count. We have represented the variations in the total magnetic moments of the studied compounds as a function of lattice constant in Figures 15 and 16. In Figure 15, it's clearly observable that the magnetic moment of $Co_2VAl$ experiences a rapid increase as the lattice constant expands from 5.2 $A^0$ to 5.4 $A^0$. Subsequently, the magnetic moment exhibits slight fluctuations within the range of lattice constants from 5.4 $A^0$ to 6.2 $A^0$.

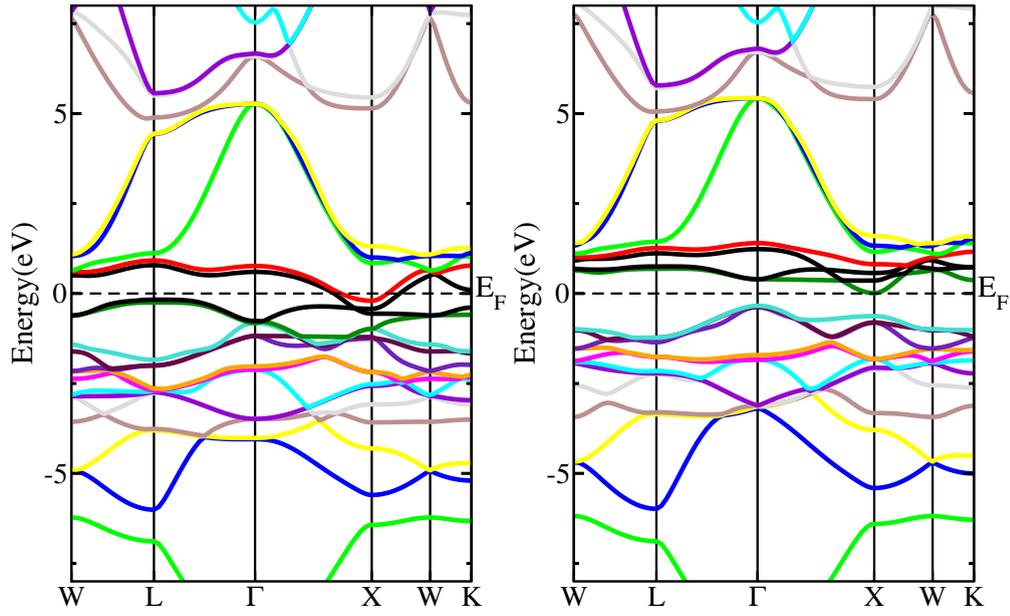

**Fig. 7** Calculated band structure of for Co$_2$VAl (spin up channel) using mbj

**Fig. 8** Calculated band structure of for Co$_2$VAl (spin down channel) using mbj

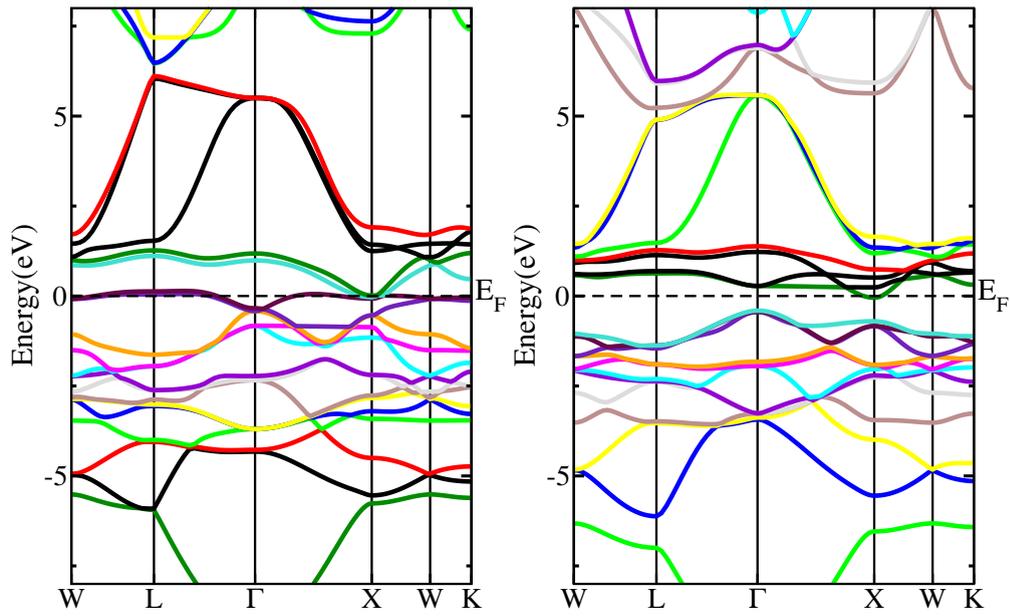

**Fig. 9** Calculated band structure of for Co$_2$VBe (spin up channel) using WCGGA

**Fig. 10** Calculated band structure of for Co$_2$VBe (spin down channel) using WCGGA

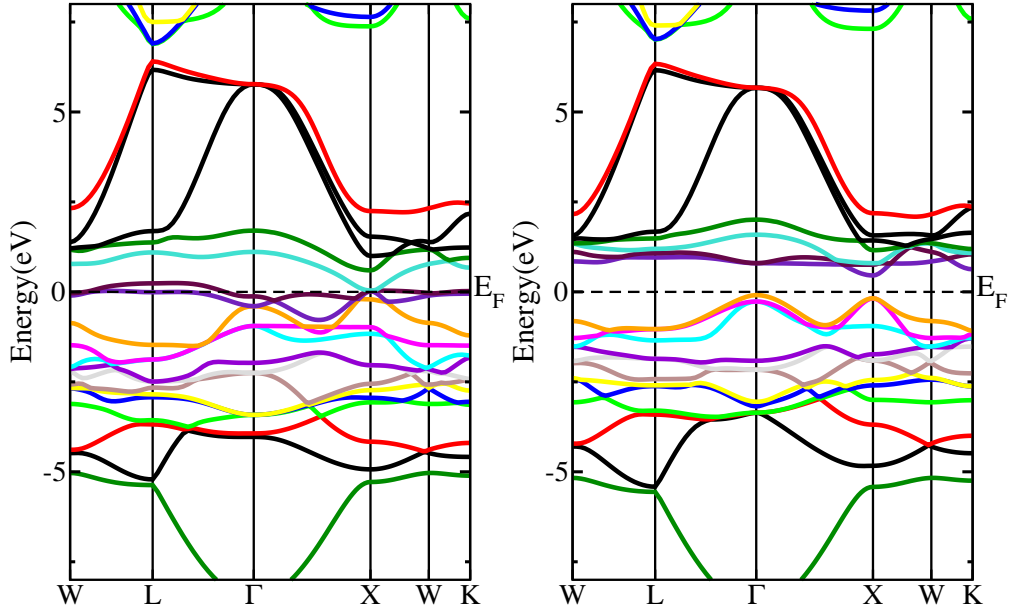

**Fig. 11** Calculated band structure of for $Co_2VBe$ (spin up channel) using mbj

**Fig. 12** Calculated band structure of for $Co_2VBe$ (spin down channel) using mbj

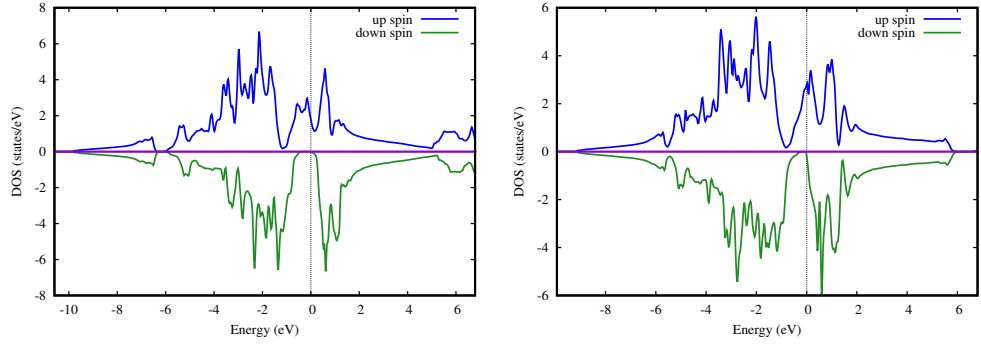

**Fig. 13** Total spin up and down DOS for $Co_2VAl$ Heusler alloy

**Fig. 14** Total spin up and down DOS for $Co_2VBe$ Heusler alloy

Figures 17 and 18 illustrate the magnetic moment variations of Cobalt and Vanadium with respect to lattice constants in $Co_2VAl$ and $Co_2VBe$, respectively. Notably, in both compounds under examination, the magnetic moment of Cobalt shows a consistent increase with lattice constant, followed by stabilization around a value approximately 1 $\mu_B$.

In contrast, the magnetic moment of Vanadium displays a decreasing trend as the lattice constant increases in both compounds, up to a lattice constant of 6.0 $A^0$. Beyond this point, from 6.0 $A^0$ to 6.2 $A^0$, the magnetic moment of Vanadium exhibits a further decrease, shifting from 0.05536 $\mu_B$ to -0.05296 $\mu_B$ in $Co_2VAl$. However, in the case of $Co_2VBe$, it undergoes an increase within the same lattice constant range, rising from -0.7087 $\mu_B$ to -0.6693 $\mu_B$.

Figures 17 and 18 showcase the magnetic moment variations of Cobalt and Vanadium concerning lattice constants in $Co_2VAl$ and $Co_2VBe$, respectively. In the case of Cobalt, it is apparent that the magnetic moment increases as the lattice constant expands and then stabilizes around a value of 1 $\mu_B$. Conversely, for Vanadium, the magnetic moment decreases as the lattice constant increases up to 6.0 $A^0$ and then experiences a slight rise from 6.0 $A^0$ to 6.2 $A^0$.

| Table 2 -: Value of Magnetic moment for $Co_2VZ(Z=Al,Be)$ | | |
|---|---|---|
| Magnetic moment per unit cell $\mu_B$ | $Co_2VAl$ | $Co_2VBe$ |
| Total | 1.958 (our work) 1.998[42] 1.55[43] 1.99[45] 1.93[47] 2.00[46] 1.981[48] 1.86 ( Exp. value)[51] | 1.000 (our work) 0.997[42] |
| Co | 1.784(our work) 1.078[42] 0.596[43] | 1.005(our work) 0.604[42] |
| V | 0.253 (our work) 0.106[42] 0.327[43] | 0.046(our work) -0.093[42] |
| Al | -0.0137(our work) -0.030[42] -0.031[43] | -0.006(our work) -0.009[42] |
| Interstitial | -0.0654(our work) -0.234[42] | -0.045(our work) -0.108[42] |

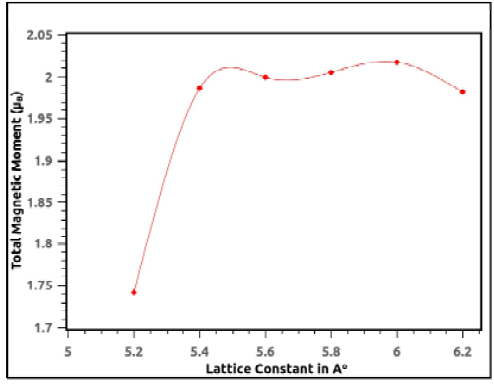 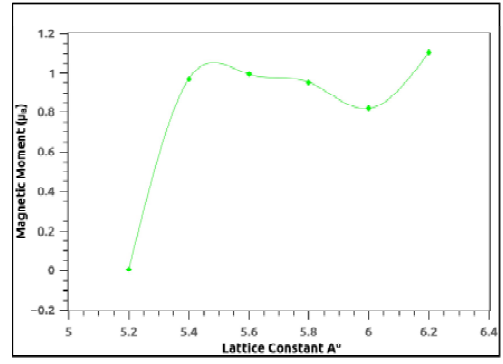

**Fig. 15** Varation of total magnetic moment with Lattice constant for $Co_2VAl$

**Fig. 16** Varation of total magnetic moment with Lattice constant for $Co_2VBe$

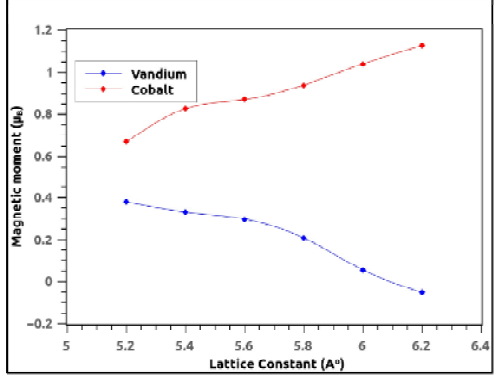 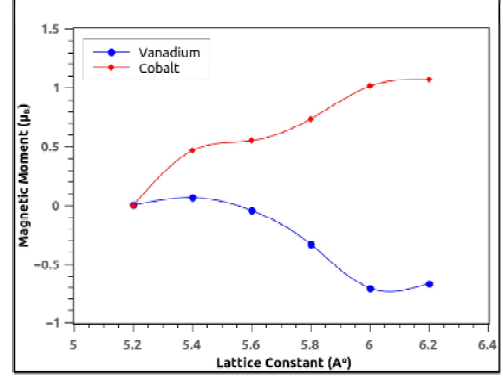

**Fig. 17** Varation of magnetic moment with Lattice constant for $Co_2VAl$

**Fig. 18** Varation of magnetic moment with Lattice constant for $Co_2VBe$

# 5 Pressure effect on structural and Electronic properties

The primary objective of this study was to investigate the impact of pressure on the structural properties and energy band gap of the material under examination. To achieve this, we employed the FP-LAPW method along with the WCGGA and mBJ approximations. By systematically varying the pressure from 0 to 30 GPa in increments of 5 GPa, we calculated the lattice constant as a function of pressure. This was achieved using the following equation [49], as given:

$$a(P) = a(0) \left[1 + \left(\frac{B'}{B}\right) P \right]^{\left(\frac{-1}{3B'}\right)} \quad (3)$$

Here, B represents the bulk modulus, B' is the derivative of the bulk modulus, and a(P) denotes the lattice parameter at pressure P. The resulting pressure-volume (P-V) data, which characterizes the relationship between pressure and volume, is depicted in Figure 17. Notably, the plot exhibits stability, indicating the absence of any structural phase transitions from the initial cubic symmetry structure to other structural phases under the range of pressures studied.

Furthermore, we examined the variation of the direct band gap as a function of hydro-static pressure for the Heusler alloys. Specifically, Figure 18 illustrates the behavior of $Co_2VAl$, while Figure 19 showcases the results for $Co_2VBe$. To understand the dependence of the semiconducting energy gap on permittivity and lattice constant, we investigated the Moss relation, Clausius Masotti relation, and quantum effects. Our findings revealed an inverse relationship between the band gap and both permittivity and lattice constant. As pressure altered these parameters, the lattice constant and permittivity responded in opposite directions. Specifically, an increase in pressure resulted in a decrease in the lattice constant, while density and subsequently permittivity increased.

As a consequence, the distinct slopes observed in the variation of the band gap with increasing pressure at different pressure ranges for $Co_2VAl$ can be attributed to the dominance of either permittivity or quantum confinement effects. In the lower pressure range (up to 20 GPa), the decreasing band gap for $Co_2VBe$ can be attributed to the influence of permittivity. However, beyond 20 GPa, both alloys exhibited an increasing band gap with rising pressure, indicating the prevalence of the quantum confinement effect at higher pressures.

The obtained data, depicting the pressure-volume (P-V) relationship, is presented in Figure 17. The plot exhibits stability, indicating the absence of any structural phase transition from the initial cubic symmetry structure to other structural phases.

Furthermore, we analyzed the variation of the direct band gap under hydrostatic pressure for the Heusler alloys. Figure 18 illustrates the trend for $Co_2VAl$, while Figure 19 showcases the results for $Co_2VBe$. The dependence of the semiconducting energy gap on permittivity and lattice constant was investigated through the Moss relation, Clausius Masotti relation, and quantum effects. The band gap was found to exhibit an inverse relationship with permittivity and lattice constant. As pressure alters these parameters, the lattice constant and permittivity change in opposite directions. Specifically, an increase in pressure leads to a decrease in the lattice constant, while density and subsequently permittivity increase.

Consequently, the distinct slopes observed in the band gap variation with increasing pressure in different pressure ranges for $Co_2VAl$ can be attributed to the dominance of either permittivity or quantum confinement. In the lower pressure range (up to 20 GPa), the decreasing band gap for $Co_2VBe$ can be attributed to the influence of permittivity. However, beyond 20 GPa, both alloys exhibit an increasing band gap with rising pressure, indicating the prevalence of the quantum confinement effect at higher pressures.

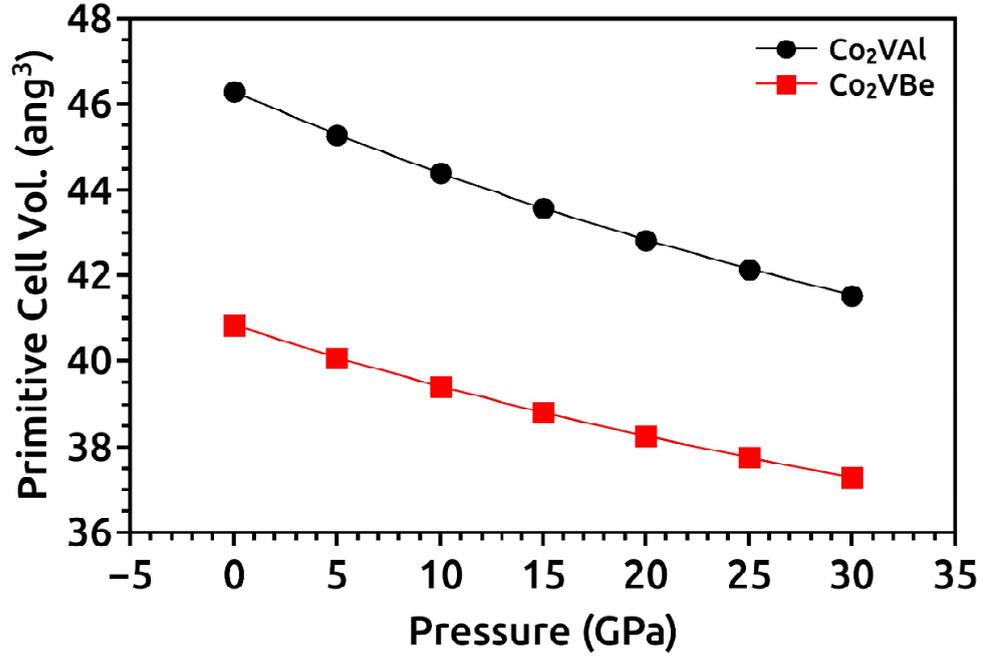

**Fig. 19** Varation of volume with applied pressure for $Co_2VAl$ and $Co_2VBe$

Table 4 -: a($A^0$) and $E_{\Gamma-\Gamma}$ as a function of pressure

| Hydro-static Pressure P (Gpa) | $Co_2VAl$ a($A^0$) | $Co_2VAl$ ($E_{\Gamma-\Gamma}$) in eV | $Co_2VBe$ a($A^0$) | $Co_2VBe$ ($E_{\Gamma-\Gamma}$) in eV |
|---|---|---|---|---|
| 0 | 5.7000 | 0.471 <br> 0.5[42] | 5.4667 | 0.461 <br> 0.41[42] |
| 5 | 5.6582 | 0.554 | 5.4325 | 0.439 |
| 10 | 5.6204 | 0.679 | 5.4018 | 0.433 |
| 15 | 5.5859 | 0.683 | 5.3739 | 0.387 |
| 20 | 5.5541 | 0.687 | 5.3484 | 0.391 |
| 25 | 5.5247 | 0.729 | 5.3248 | 0.423 |
| 30 | 5.4974 | 0.792 | 5.3030 | 0.451 |

# 6 Conclusion

In this research, we conducted a comprehensive investigation of the energy band gap in $Co_2VZ$ (Z = Al, Be) Heusler alloys under varying high pressure conditions (P = 0.0, 5.0, 10.0, 15.0, 20.0, 25.0, 30.0 GPa) using the mBJ approximations. One important finding of this study is that the pressure-volume (P-V) curve exhibited stability, indicating the absence of any structural phase transition within the investigated pressure range. This suggests that the crystal structure of $Co_2VZ$ (Z = Al, Be) Heusler alloys remained unchanged under the applied high pressures. Furthermore, we observed different slopes in the variation of the band gap with increasing pressure at different pressure ranges for $Co_2VZ$ (Z = Al, Be) Heusler alloys. These distinct trends can be attributed to the dominance of either the permittivity or the quantum confinement effects. The interplay between these factors influences the behavior of the band gap as pressure increases. By comparing our results with earlier studies that employed different approximations, we contribute to the body of knowledge and provide valuable insights into the energy band gap behavior of $Co_2VZ$ (Z = Al, Be) Heusler alloys under high-pressure conditions. This research enhances our understanding of the fundamental properties of these materials and their potential applications in various fields, including advanced electronics, spintronics, and energy storage.

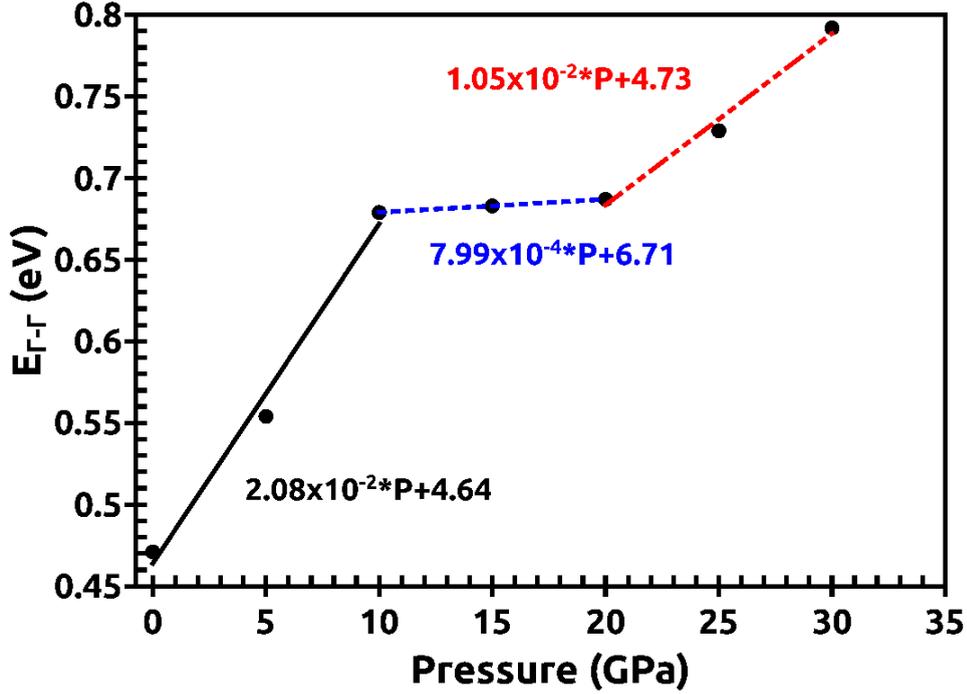

**Fig. 20** Band gap as a function of Pressure for $Co_2VAl$ Heusler alloy

# 7 Credit authorship contribution statement

(**S.Kumar, Diwaker, Karan. S. Vinayak**): Software, Workstation, Conceptualization,

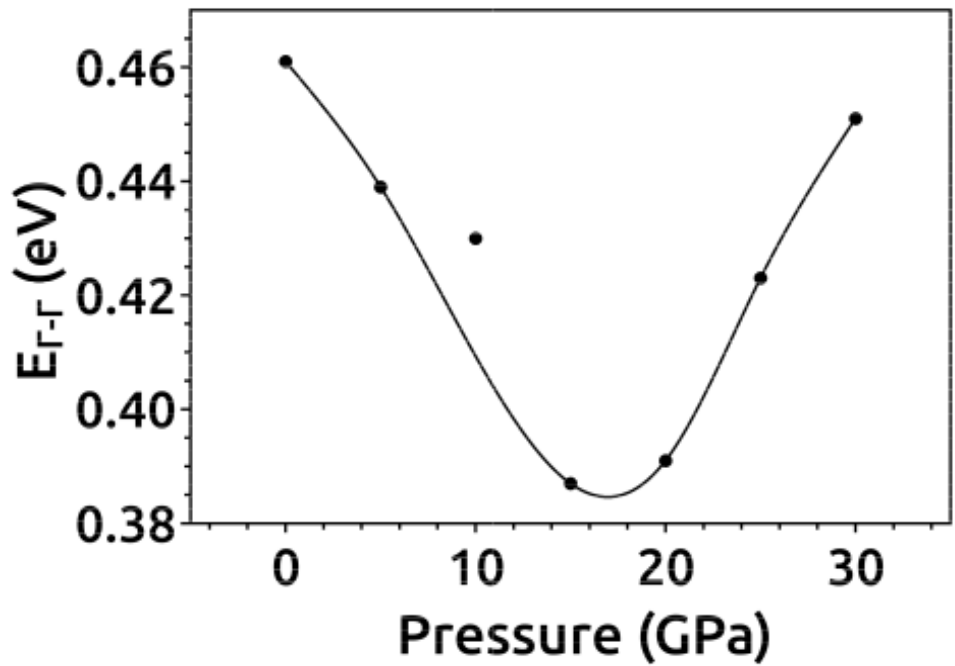

**Fig. 21** Band gap as a function of Pressure for $Co_2VBe$ Heusler alloy

(**S.Kumar, S. L. Gupta, Diwaker**)-: Methodology, Data curation, Software handling, Writing - original draft Writing - review and editing.

# 8 Declaration of Competing Interest

The authors declare that they have no known competing financial interests or personal relationships that could have appeared to influence the work reported in this paper.